\documentclass[10pt,letterpaper,twocolumn]{hotnets15}
\usepackage[latin9]{inputenc}
\usepackage{color}
\usepackage{verbatim}
\usepackage{amssymb}
\usepackage{graphicx}
\usepackage[unicode=true,pdfusetitle,
 bookmarks=true,bookmarksnumbered=false,bookmarksopen=false,
 breaklinks=false,pdfborder={0 0 0},backref=false,colorlinks=true]
 {hyperref}
\usepackage{breakurl}

\usepackage{listings}
\makeatletter

\@ifundefined{date}{}{\date{}}
\usepackage{epsfig}
\usepackage{endnotes}

\usepackage{enumitem}
\setitemize{noitemsep,topsep=0pt,parsep=0pt,partopsep=0pt,leftmargin=*}

\usepackage{times}
\usepackage{url}
\usepackage[font=small,format=plain,labelfont=bf,up,textfont=it,up]{caption}

\usepackage{fancyhdr}

\usepackage{color}
\usepackage{balance}

\usepackage{hyperref}
\hypersetup{pdfstartview=FitH,pdfpagelayout=SinglePage}
\newcommand{\yahoo}{Yahoo}
\newcommand{\yperf}{YTrace}
\newcommand{\ytrace}{YTrace}
\newcommand{\sherpa}{Sherpa}
\newcommand{\hide}[1]{#1}

\makeatother

\begin{document}

\title{YTrace: End-to-end Performance Diagnosis in\\
Large Cloud and Content Providers}

\newcommand{\supsym}[1]{\raisebox{4pt}{{\footnotesize #1}}}
\newcommand{\gt}{\supsym{$\dag$}}
\newcommand{\tf}{\supsym{$\ddag$}}
\newcommand{\cm}{\supsym{$\S$}}
\newcommand{\pu}{\supsym{$\bot$}}

\author{Partha Kanuparthy\tf, Yuchen Dai\gt, Sudhir Pathak\gt, Sambit Samal\gt,\\Theophilus Benson\cm, Mojgan Ghasemi\pu, P. P. S. Narayan\gt\\
\affaddr{\tf~Yahoo Research~~~~ \gt~Yahoo Inc.~~~~ \cm~Duke University~~~~ \pu~Princeton University}}

\maketitle

\begin{abstract}
Content providers build serving stacks to deliver content to users.
An important goal of a content provider is to ensure \emph{good user
experience}, since user experience has an impact on revenue. In this
paper, we describe a system at \yahoo{} called \yperf{} that diagnoses
bad user experience in near real time. We present the different components
of \yperf{} for end-to-end multi-layer diagnosis (instrumentation, methods
and backend system), and the system architecture for delivering
diagnosis in near real time across all user sessions at \yahoo{}.
\yperf{} diagnoses problems across service and network layers in the
end-to-end path spanning user host, Internet, CDN and the datacenters,
and has three diagnosis goals: detection, localization and root cause 
analysis (including cascading problems) of performance problems 
in user sessions with the cloud. The key component of the methods 
in \yperf{} is capturing and discovering
causality, which we design based on a mix of instrumentation API,
domain knowledge and blackbox methods.
We show three case studies from production that span a large-scale
distributed storage system, a datacenter-wide network, and an end-to-end
video serving stack at \yahoo{}. We end by listing a number of open
directions for performance diagnosis in cloud and content providers.
\end{abstract}

\section{Introduction}

\begin{figure}
\begin{centering}
\includegraphics[clip,width=0.9\columnwidth]{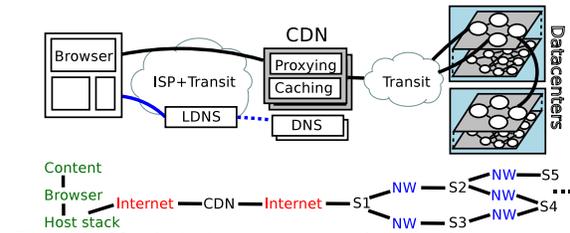}
\par\end{centering}

\protect\caption{Model of a large content provider showing the end-to-end path for
a user session.\label{fig:Model-CP} Lower figure shows canonical
execution graph to determine instrumentation; ``S'' and ``NW''
represent services and network respectively.\label{fig:Canonical-execution-tree}}
\end{figure}

Large content providers \hide{such as Yahoo, Google, Netflix and Facebook}{}
serve users from large-scale serving stacks in geographically distributed
datacenters on the Internet. They can be modeled as cloud infrastructure
that consists of multiple datacenters and a Content Distribution Network
(CDN) (Figure \ref{fig:Model-CP}). Users interact with the content provider
by making RPCs (also called \emph{user sessions}) to the CDN and the 
datacenters. The \emph{user experience} of
a user session with the provider depends on several factors from the
serving stack, to the datacenter network and the Internet, to the content. \emph{Bad} user experiences
result in loss of users and revenue \cite{websiteperf}.

Content providers build for good user experience by building high-performance
serving stacks and network infrastructure. Serving stacks are compositions 
of services, and services are usually
large distributed systems comprising of hundreds to thousands of
hosts -- on top of the datacenter network and inter-datacenter wide area
paths. Serving stacks include latency-tolerant distributed execution techniques
such as parallelism and redundancy \cite{dean2013tail}. For example,
a user request for a personalized web page could be served by ``assembling''
parts of the page, each generated by a service\footnote{Such designs are also called 
service-oriented and microservices architectures.}. In order to do this,
services (specifically, hosts in a service) make RPCs to each other
over the underlying network paths.

Due to the composition scale and heterogeneity of a serving stack, it is prone
to performance problems that span multiple layers -- from the infrastructure
layer such as network and servers, to the higher layers such as the
OS, containers and service processes within a server, to the distributed
systems layer -- and localized among nodes in the end-to-end
path (Figure \ref{fig:Model-CP}). Detecting and troubleshooting bad
user experience is a complex and tedious problem at scale, since it often involves
multiple services and layers, and hence, coordination between multiple teams across service tiers
and underlying layers. It is hence equally important to build systems
that continuously monitor and diagnose bad user experiences. Such
systems help troubleshoot to quickly fix performance problems, and
know where to allocate resources in the medium-term. Further, near real time
\emph{diagnosis as a service} is a useful primitive to optimize
existing systems against performance problems. Content providers
have designed and deployed several systems in practice \cite{chow2014mystery,kim2013root,krishnan2009moving,ren2010google,sigelman2010dapper,sun2011identifying,yu2011profiling,zhu2012latlong};
however, these systems do not diagnose performance problems end-to-end
and across layers.

\begin{figure*}
\begin{centering}
\includegraphics[width=0.9\textwidth]{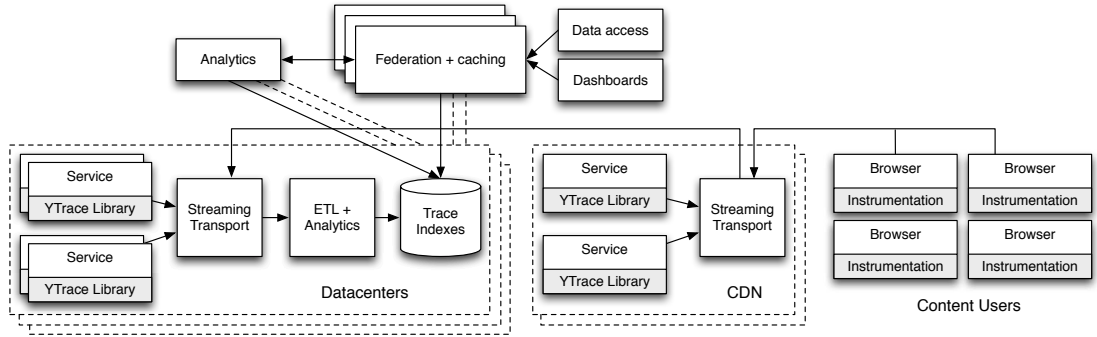}
\par\end{centering}

\protect\caption{\yperf{} architecture and components.\label{fig:YPerf-architecture}}
\end{figure*}

We present \yperf{}, a system that we are building at \yahoo{} to
diagnose end-to-end performance problems that impact user experience.
\yperf{} has three components: instrumentation to collect data, diagnosis
methods that run on the data, and an efficient backend to index the
data and execute diagnosis queries (Figure \ref{fig:YPerf-architecture}).
In this paper, we focus on the first two components and touch upon the 
third. We consider dynamic
web content that is tailored for users -- perhaps the most common
on the Internet. Our definition of user experience depends on the content type:
for web content, we estimate user experience as the 
page load time -- the latency between the user's content request
and the Javascript OnLoad event; and for video streams, we consider 
duration of rebuffering events. Our work can easily be extended to
diagnose performance problems with other content types and definitions
of user experience.

When building an end-to-end diagnosis system, there are key requirements
for large content providers:
\begin{itemize}
\item Tie to user experience: Instrumentation and diagnoses should directly
relate to user experience of real users.
\item The diagnosis output should be general enough to help troubleshoot
\emph{almost all} performance problems, including casding failures.
\item Multi-layer: The diagnosis should span as many layers in the serving
stack as possible. At a minimum, it should include all services, the
host machines and the underlying network layer.
\item Instrumentation should have low overhead, so it does not affect the
user experience.
\item Accuracy: Diagnosis should have low false positive and false negative
rates for the use cases. It should be able to diagnose tail latency.
\end{itemize}

The key ideas behind \yperf{} rely on identifying concurrent event execution,
both at the service level and in the network. Knowing the context of concurrency
enables \yperf{} to compute the most important information for diagnoses -- the
critical path in the execution. In order to find concurrency, \yperf{}
records and mines causal relationships between events in a user session at
the service and network layers. It aggregates diagnoses across user sessions
and renders an interactive dashboard geared towards troubleshooting.

\textcolor{red}{}%

\section{Problem Statement}\label{sec:Formulation}

There are three broad classes of use cases of \yperf{}: troubleshooting,
resource provisioning and service adaptation. Troubleshooting aims
to fix performance problems that users face after the problems occur.
It requires the system to deliver near real time, actionable, insights
into performance problems. Resource provisioning is a relatively longer-term
task that involves querying the system for aggregate views of diagnosis
to find where to add resources\footnote{Resource provisioning also requires answers to ``what-if'' questions.
This is outside the scope of our current work.}. Service adaptation uses \yperf{} as
a near real time diagnosis-as-a-service to build serving stacks that optimize for
user experience. For example, the traffic engineering service
at a CDN may route users to CDN nodes based on diagnoses of Internet
paths; the rate adaption module in a video player may make strategic rate
choices if it had diagnoses. Since this involves pre-defined queries, the system may materialize
such queries to minimize query times\footnote{Large query delays can be detrimental
to performance, e.g., in load balancing \cite{mitzenmacher2000useful}.}.

Based on discussions with teams across Yahoo, we formulate a problem 
statement whose solution provides actionable
input for the three use cases. \yperf{} has three goals for every
user session:
\begin{description}
\item [{Detection:}] Is a user session seeing a performance problem?
\item [{Localization:}] Where are the performance problems in the end-to-end
path (and across all layers)?
\item [{Root~cause~analysis:}] Why are the performance problems occurring?
\end{description}

In addition to per-session diagnoses, the \yperf{} backend 
supports (and materializes views of) aggregate queries over multiple user sessions, such as
clusters of users (e.g., ISP and geography), and over a service
in the datacenter in a time window. Aggregate queries with such predicates
enable statistically significant analyses while conditioning on confounding
variables.

\section{Instrumentation}\label{sec:Instrumentation-Space}

The first step towards performance diagnosis of a user session
is instrumentation of components that participate in the session. The
instrumentation should not add significant latency to the session.
The key is to determine \emph{necessary and sufficient} instrumentation for diagnosis.
We implement optimized libraries for
instrumentation so that the instrumentation overhead 
is very low relative to end-to-end latency.

One way to determine instrumentation is by considering the canonical
end-to-end user session graph, whose nodes are
components (which impact user experience) that participate in user sessions and whose edges represent
point-to-point communication between nodes; it should cover all components and layers
that are necessary for diagnosis. Figure \ref{fig:Canonical-execution-tree}
shows the graph  for large content providers that spans: the user 
end-host, the CDN, the serving stack
spanning one or more datacenters, and the underlying network infrastructure.

In order to diagnose performance problems with each
node in the canonical graph, the necessary and sufficient instrumentation
will include performance data from every node in the
graph (necessary condition) and will not include redundant instrumentation
between edges (sufficient). The necessary and sufficient instrumentation
for root cause analysis at a node depends on the attributes \yperf{}
needs to be able to fingerprint and match problem signatures (\S\ref{sec:Diagnosis}).

\yperf{} includes two forms of instrumentation: (1) synchronous instrumentation
that is in-band with the user session, and (2) asynchronous instrumentation 
from components that cannot be modified for instrumentation (such as network
devices). We implement synchronous instrumentation in the form of 
\emph{distributed tracing}, which allows \yperf{} to tie performance of any
component into the user experience. \yperf{} uses causal relationships in
instrumentation data to diagnose performance problems.


\subsection{Synchronous Instrumentation}

\noindent\textbf{User-side instrumentation.} User end-host
instrumentation enables \yperf{} to diagnose performance problems
with events at the end-host (includes browser, any containers, and the content itself).
In general, content providers cannot alter the browser (e.g., by introducing
plugins), which leaves them with a limited set of user-side performance
measurements.

The work of content in a browser can be modeled as a sequence of events
spanning fetching resources (either via local cache or network), execution
and rendering. The W3C Navigation Timing (NavTiming) recommendation
\cite{navtiming} describes such an event model for origin content
(the page HTML) and exposes it via a Javascript API to the web page.\footnote{A similar event API for the other resources that the page requires
is supported by the W3C Resource Timing recommendation \cite{restiming}.}

We use the NavTiming event model for user-side instrumentation for web content in
\yperf{}. This enables us to break down the user experience (page
load time) into timing of events for origin content. There is a causal
relationship between all NavTiming events: for example, DNS lookup (if any)
causes TCP connect to the CDN. In particular, all events measured by 
NavTiming and Resource Timing have well-defined causal relationships.
Having causal relationships helps us understand the events that resulted
in bad user experience, since they are necessary to construct the 
latency critical path of the session.

In the case of video streaming (which is a linearizable sequence
of RPCs per-segment to the CDN), \yperf{} uses an event model that includes timing
of per-segment events at the video player. These events span
segment RPCs, and decoding and rendering of those segments to the screen.

\textcolor{red}{}
\begin{figure}
\begin{centering}
\includegraphics[bb=0bp 70bp 683bp 312bp,clip,width=1\columnwidth]{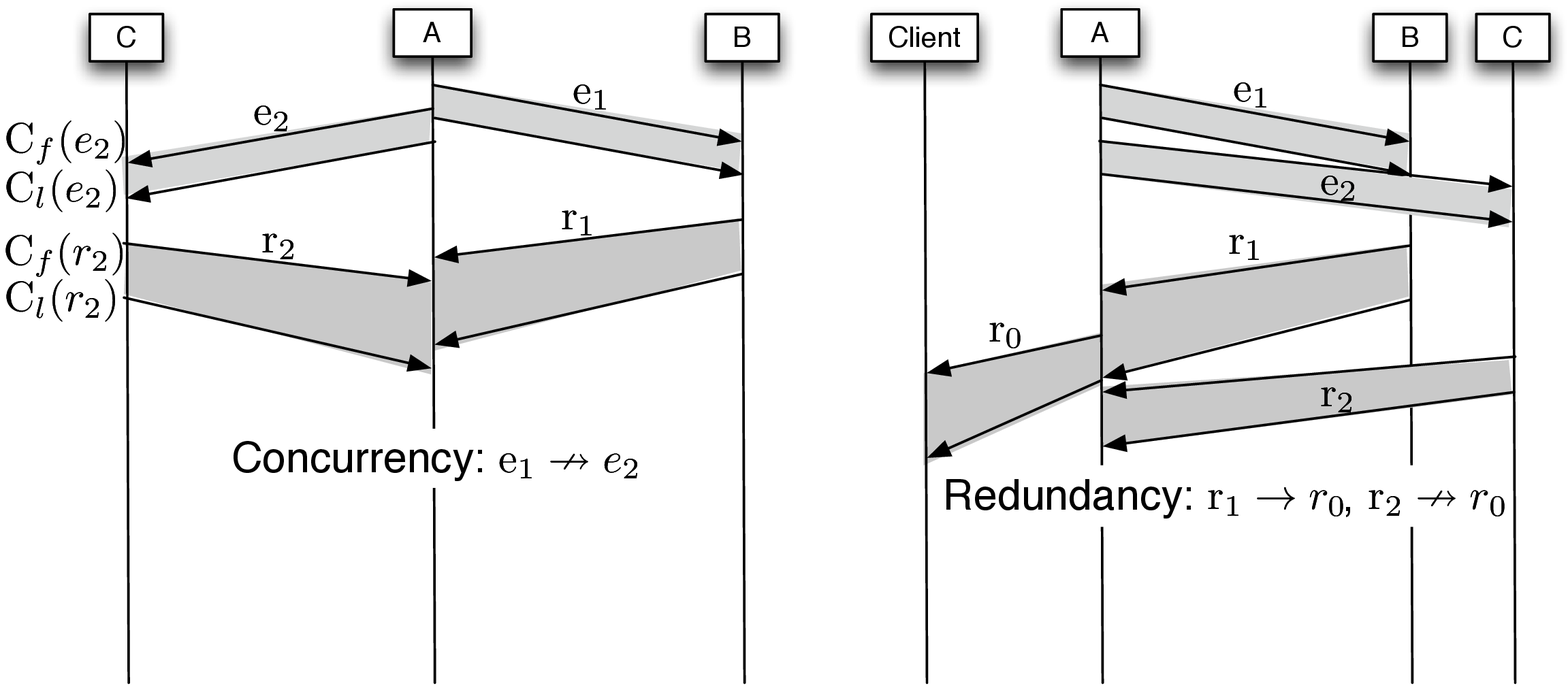}
\par\end{centering}

\textcolor{red}{\protect\caption{Inferring causality in RPC patterns.\label{fig:Inferring-causality}}
}
\end{figure}

\vspace{0.1in}\noindent\textbf{Distributed tracing.}
\yperf{} synchronously traces user sessions (i.e., RPCs from user agent) through all
execution nodes in the serving stack, including the CDN services and the user host
(see user-side instrumentation above). Distributed tracing is a
common monitoring primitive in large-scale serving stacks~\cite{chow2014mystery,sigelman2010dapper},
and involves two steps: assigning a globally unique ID to a user session, and
propagating this ID through all nodes in the serving stack. The ID propagation
is typically implemented by adding the ID to all RPCs during session execution
-- for example, in the form of a serialized header. \yperf{} records the timing
of events related to each RPC during session execution using node-local clocks.

When implementing distributed tracing, there are a few design constraints
that arise from large-scale environments. Such environments are highly
heterogeneous, not only in the platforms used, but also in runtime complexity
such as RPC execution patterns (see Figure \ref{fig:Inferring-causality}), 
serialization formats and protocols. We find two forms of RPC-level concurrency
in distributed execution: parallelism and redundancy -- both in the context of
RPC ``fanout'' implementations. Parallelism includes parallel RPCs; the opposite of
which is serialized RPC execution. Redundancy is a case where a service doing
the fanout only waits for a few responses before sending back a response to the caller;
this is typically used in search engine stacks.

Perhaps the most important requirement in distributed tracing implementations is to
\emph{capture causality} among \emph{all} events in the end-to-end execution. Ideally,
causality among events should be described by the services themselves (during tracing),
and not inferred offline using tracing data (an approach adopted by 
prior work \cite{aguilera2003performance,chow2014mystery}). 
The reason for this is dynamic
behavior in web services: for example, the concurrent/serialized and redundant 
RPC execution patterns shown in Figure~\ref{fig:Inferring-causality} can be triggered
as functions of session attributes, performance history and runtime environment. Such
dynamicity makes offline inference of causal relationships between RPC edges hard
(without additional data that may not be within the scope of a tracing system).

\ytrace{} captures causality between RPC ``edges'' (requests and responses),
since causality in RPC execution patterns such as concurrency and redundancy
exists between RPC edges. This RPC model is a key difference between \ytrace{}
and prior tracing systems such as Dapper (and Zipkin), which capture causality
at the granularity of RPCs. \ytrace{} captures causality in two forms: during
tracing using service-level APIs designed to capture causality, and offline using
(well-defined) happens-before relationships~\cite{lamport1978time}.

Services call the \ytrace{} tracing API at each RPC edge. The API returns an immutable
session context (passed as a handle) for each incoming RPC, until that RPC is fully served. The
API times each RPC edge, and any annotations\footnote{The annotations are optional service-specific
timestamped key-value pairs, such as lock events. They are used by developers to
understand service-specific performance, such as impact of lock contention on
end-to-end performance.} across the session. It consumes 
and returns all headers that are/should be serialized in RPCs.
\begin{lstlisting}
void *handle = create(/*String*/ in_header);
String out_header = sendtonext(handle);
recvfromnext(handle, /*String*/ in_header);
String out_header = sendtoprev(handle);
annotate(handle, /*String*/key, /*String*/val);
close(handle);
\end{lstlisting}
The API captures causality in two ways. First, the session context (handle)
allows \ytrace{} to capture parent-child relationships between RPC edges.
Second, the headers that the API generates include an RPC ID and parent RPC ID
(that are unique to the session), and these IDs record
causality between the request and response(s) (if any) of an RPC. The parent-child
RPC IDs also record global ordering of RPCs in the session. 

We implement the \ytrace{} tracing API as a userspace library
that services call during execution. The library provides a 99th percentile runtime SLA of
3$\mu s$ per RPC. The low runtime overhead of the library is due to two reasons.
First, the APIs are stateless, since the session state (handle) in a 
service is immutable. The state, instead, is passed over the network
in the serialized headers; an example of such state is the child-parent
RPC IDs in the session. This design choice trades-off expensive state 
maintenance in the API with a few additional bytes on the network; and
also avoids any need for synchronization at session-level in a service.
Second, all logging in \ytrace{} is asynchronous.

The API captures a significant set of causal relationships in sessions,
but it does not capture causality (or lack of it) in RPC execution patterns such as parallelism and redundancy in
the execution graph \cite{chow2014mystery,dean2013tail}; see Figure
\ref{fig:Inferring-causality} -- we found that doing so makes the API
complex (which slows down adoption). \yperf{} uses happens-before 
relationships to infer this causality (see $\S$\ref{sec:Diagnosis}).

\begin{figure}
\begin{centering}
\includegraphics[width=1\columnwidth]{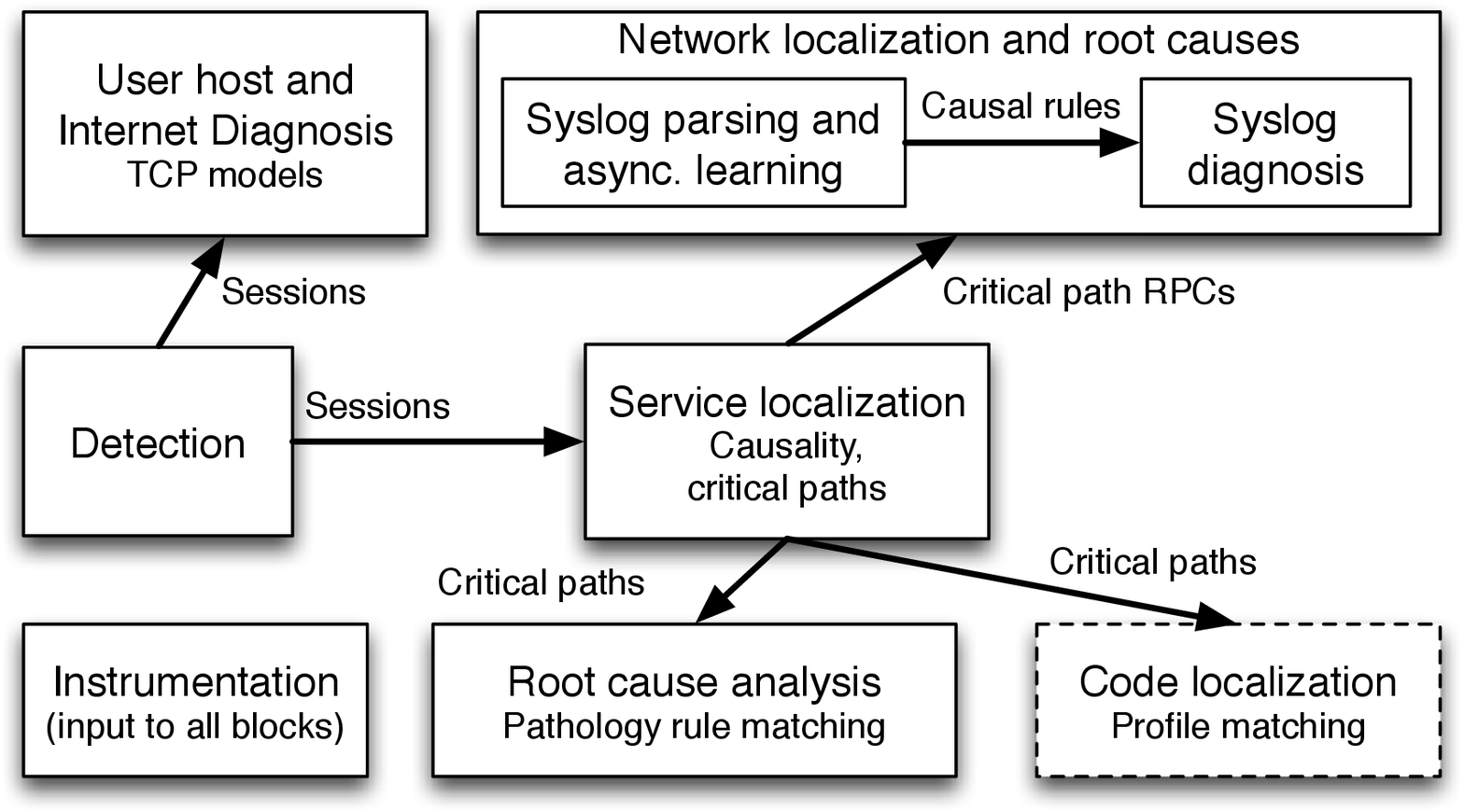}
\par\end{centering}

\protect\caption{\yperf{} diagnosis flow.\label{fig:flowchart}}
\end{figure}

\vspace{0.1in}\noindent\textbf{CDN instrumentation.} 
The wide area Internet path can be a significant source of
performance problems that impact user experience.
In order to diagnose these problems and their
impact on user experience, we need to instrument the Internet path in isolation.
Typical approaches explored in literature include active probing 
(which is asynchronous) or model-based methods that infer path 
performance user-side and/or CDN-side measurements of the trace.
Both approaches have limitations: the former adds network traffic
and may not be causally related to the session (since it is asynchronous), 
while the latter methods are prone to user host problems (which are
not uncommon).

In order to diagnose and isolate Internet performance problems,
we take periodic snapshots of measurements that the TCP stack in
the CDN kernel maintains, for the TCP connection used by user host
RPCs. Specifically, we snapshot \texttt{tcp\_info} structures
from the Linux kernel. The structure contains end-to-end
statistics of the TCP connection that affect serving performance,
such as packet retransmissions, reordering, RTT, sender and receiver
windows, etc.; it hence measures the Internet path as the flow sampled
it. \yperf{} uses the TCP connection statistics to localize throughput
bottlenecks to sender (content generation), receiver and path-based
limitations; and to diagnose download bottlenecks with user hosts. 
Note that we do not include diagnosis of CDN traffic
engineering-related performance problems; in other words, \yperf{}'s
diagnosis is conditioned on the traffic engineering decision for a
user session (see discussion, \S\ref{sec:Discussion-and-Conclusion}).

\vspace{0.1in}\noindent\textbf{Process profiling.} We are adopting
Continuous Profiling \cite{anderson1997continuous,ren2010google}
in \yahoo{} services for a small fraction of user sessions. At a
high level, Continuous Profiling collects performance counters exposed
by modern CPUs. Performance counters allow us to understand host-level
bottlenecks and localize bad user experience down to code using the 
associated program counters (in conjuction with the process binaries).

At the end of each session, \ytrace{} records a directed trace graph that 
includes: \emph{(i)} the end-to-end execution graph with compute,
serialization and RPC timings at each node and event causality, \emph{(ii)} user-side event
timings and event causality, and \emph{(iii)} TCP-layer measurements of
user-side Internet at the CDN.

\subsection{Asynchronous Instrumentation}

Despite service stack-level concurrency, the underlying network is a shared resource
and typically underprovisioned (e.g., fat-tree datacenter topologies).
The network can introduce performance problems in RPCs during session
execution, which can impact user experience.
\yperf{} collects asnchronous instrumentation from components in the
end-to-end serving stack that cannot be modified to do tracing. Such
components are typically in the underlying network layer, such as the
datacenter network devices.

\yperf{} collects \emph{syslogs} from all datacenter network devices.
Syslogs include detailed and fine-grained state information of each
device and the root cause. In conjunction with syslogs, \yperf{} uses 
network topology to localize user session performance
problems to network devices. It collects network topology snapshots
of: \emph{(i)} the wide area Internet paths from the CDN to \emph{client
clusters} and to the datacenters using traceroutes, and \emph{(ii)}
each datacenter network using device configurations.

\section{Diagnosis}\label{sec:Diagnosis}

\yperf{} uses synchronous and asynchronous instrumentation to diagnose performance problems
that impact user experience. In this section, we sketch the diagnosis
methods. See Figure \ref{fig:flowchart} for a flow overview.

\vspace{0.1in}\noindent\textbf{Detection.} The first step towards
performance diagnosis is to detect performance problems. Since our focus is on user experience, we frame
the detection problem around it: Is the page load time\footnote{We can ask similar detection questions about user experience metrics
for other content types, e.g., rebuffering in video.} \emph{large} for the user session? \yperf{} answers this question
by \emph{estimating a baseline }(normal behavior) for the page load
time based on history, and finding if the session has a statistically
significant deviation from the baseline. The page load time is measured
at the user-side. Note that detection algorithms have to be aware
of confounding variables such as the web page (session) attributes, the user attributes and
time of day; \yperf{} conditions the detection based on domain knowledge
of pre-defined confounding variables.
\yperf{} also supports detection based on other definitions of user
experience, or not based on user experience. For example, video quality of experience,
abnormal service latencies or unusual execution graphs for a session.
\yperf{} currently estimates a simple baseline as the historic inter-quartile range,
since we are interested in understanding performance behind both low and high
user experience metrics.

\subsection{Content Diagnosis}

For sessions that were detected as having performance problems, 
\yperf{} uses the user-side instrumentation to determine whether there were performance
problems that were localized to the user agent (browser). To do this,
it checks whether the latency of user-side events from the Navigation
Timing API \cite{navtiming} (e.g., DOM processing and rendering)
are significant relative to the OnLoad time for the page. Note that
NavTiming events are causally related and the critical path of user-side
events includes all events.

In general, resources on the page are fetched (and executed) concurrently
with the origin page -- and the concurrency depends on the origin
content, ordering of resource arrivals and execution latency (parsing,
DOM construction, etc.). This dependency between resources leads to
blocking periods; however, such analysis requires browser modifications
\cite{wang2013demystifying}. \yperf{} currently treats all content
(origin and resources) performance as independent (note that 
origin content is typically the dynamic, non-cacheable
content in personalized pages). We are investigating dependency and
blocking time measurement that avoids browser changes.

\subsection{Service Diagnosis}

Service localization refers to the question of which services in the session execution
graph \emph{caused} a performance problem for sessions that were
detected as having performance problems. \yperf{} localizes service-level
problems by estimating the critical path(s) -- in terms of service
latencies -- in the session execution graph. In order to compute the
critical path, \yperf{} needs context of concurrency in the
execution graph, which is determined by causality between RPC edges.

\yperf{} tracks causality as follows. The synchronous instrumentation
system tracks two forms of RPC causality during tracing:
parent-child RPC relationships, and request-response causality.
In order to keep the tracing API simple to use and reduce usage errors, we do not track
(at the API-level) causality \emph{between} sibling RPC edges at a single node. Causality
between sibling RPC edges may (or may not) exist, depending on the
RPC execution patterns used (see Figure~\ref{fig:Inferring-causality}). 
Since such patterns are typically dynamic, based on session and environment
attributes, we cannot use blackbox methods that mine causal relationships
from offline session trace data \cite{aguilera2003performance,chow2014mystery}.

The parallelism and redundancy RPC patterns lend well to happens-before 
relationships (directly follows from their definition).
Consider outgoing RPC edges $e_{1}\dots e_{n}$ at a service node $A$ (Figure \ref{fig:Inferring-causality});
and let the responses to the outgoing edges be the edges $r_{1}\dots r_{n}$
(incident on $A$). Denote the first byte and last byte timestamps ($A$'s
wall clock) of an edge $e$ by $C_{f}(e)$ and $C_{l}(e)$. Timestamps
are taken from user space. Note that $e_{i}$ causes $r_{i}$ for all $i$,
denoted as $e_{i}\rightarrow r_{i}\forall i$.\\
\noindent Causality in \emph{non-parallelism}: $r_{i}\rightarrow e_{j}$:
$C_{f}(r_{i})<C_{f}(e_{j})$; $r_{i}\nrightarrow e_{j}$ otherwise.\\
\noindent\emph{Redundancy:} If edge $r_{0}$ is $A$'s reply to calling
node, $r_{i}\rightarrow r_{0}$ $\forall$ $r_{i}$: $C_{f}(r_{i})<C_{l}(r_{0})$;
$r_{i}\nrightarrow r_{0}$ otherwise.

\yperf{} uses edge causality to estimate
the critical path in the execution graph, defined as the \emph{causal} round trip path in
the graph with the largest total (service and network) latency.
Latency at a service with an incoming edge $e_{0}$ and a causal outgoing
edge $e_{1}$ is the computation time: $l_{01}^{s}=C_{f}(e_{1})-C_{l}(e_{0})$.
The network latency is the RPC edge (de)serialization time at the caller node. Note that the
critical path in the same execution graph can change based on RPC
causality: for example, if service $A$ makes two RPCs $e_{1}$ and
$e_{2}$ to $B$, the critical path may include one or both of $e_{1}$
and $e_{2}$ depending on $e_{1}-e_{2}$ causality:\\
\includegraphics[clip,width=0.8\columnwidth]{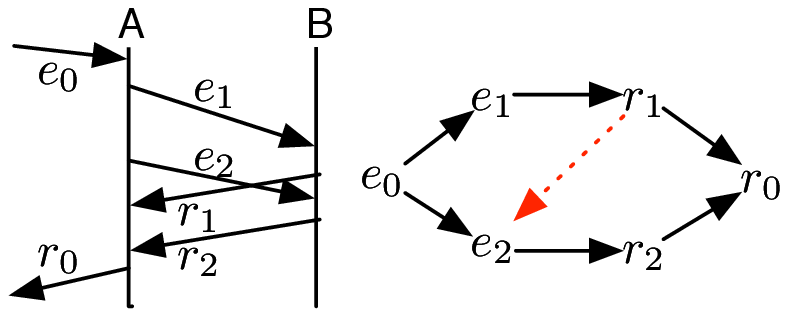}

\yperf{} estimates the contribution of a service as the sum of computation
latencies for all incoming-causal outgoing edge pairs $l_{ij}$. It
reports the service-level localization output as the top service contributors,
and their fraction of end-to-end (user-side) latency, amongst services in
the critical path.

\subsection{Network Diagnosis} 
\label{sec:netdiag}

\yperf{} uses syslogs from network devices to diagnose datacenter network
problems. It uses TCP stack measurements at the CDN to isolate problems
on the user-to-CDN Internet path (note that we cannot instrument the Internet path).
We first look at datacenter network problem diagnosis.

\noindent\textbf{Datacenter network diagnosis.}
The critical path found during service localization represents the subset of execution
that contributed to user latency, and it includes time spent by
RPCs in the network. The network can degrade the performance of RPCs
by inducing latency, packet losses and reordering, which increase
the RPC time and reduce throughput (especially for RPCs with large
payloads). Our goal in network diagnosis is to localize and find
root causes of datacenter network problems. We are interested in
localizing cascading problems and finding root causes that propagate
across the network stack; for example, a hardware problem in a switch
that cascades into problems in the connected router as both L2 and routing
plane problems.

\yperf{} uses syslogs and the datacenter network topology to diagnose
cascading problems. Each datacenter network device emits a stream of
syslog messages, which are semi-structured text that include a timestamp,
severity level and semantics of the problem (network interface, problem
type and attributes, etc.). Our goal is to represent a problem as a
structured graph that describes the causal activity (the cascade) in the problem.
It uses domain knowledge to preprocess syslogs: mapping them to structured
``templates'' (including equivalence classes of problem types), and 
extracting device attributes (if any). We leverage some of the prior
work on template extraction \cite{Qiu:2010:HMN:1879141.1879202}. 
The domain knowledge is a one-time
input to \yperf{} and does not need changes unless the syslog templates
change (e.g., due to vendor or major OS changes, which are infrequent).

The first step towards diagnosing session performance due to network problems
is to find RPCs in the session critical path that impacted user experience.
\yperf{} computes a candidate list of RPCs per-session as follows.
Consider two services $A$ and $B$, and an RPC $e$ from $A$. The local clock
at node $A$ is $C_{f}^{A}$ (we consider timestamps at the start
of each RPC request/response to avoid self-loading effects). For diagnosis,
we are interested in the variable component of the one-way delay
-- typically queueing delays. \yperf{} estimates queueing delay during
the RPC as $d_{AB}(e)=C_{f}^{B}(e)-C_{f}^{A}(e)-\min_{\Delta}(d_{AB})$,
where the $\min$ is taken over the recent $\Delta$ time window\footnote{The time window should be large to include queue dissipation ($\mu s$
in datacenter networks) but smaller than clock drift timescales (mins.).}; the $\min$ is an estimate of the constant components of the one-way
delay. \yperf{} detects an RPC as having a performance problem if the 
queueing delay is significant relative to the end-to-end latency.
It then computes a set of network devices that the RPC could have traversed.

The typical way to localize network-level performance problems is boolean
network tomography \cite{Duffield:2003:SNP:948205.948232} on end-to-end
observations of RPC queueing delays. Tomography takes as input observations
of paths -- good and bad -- that \emph{overlap} with a problem path.
It aims to isolate the part of the network that led to the bad observations.
Tomography is not directly applicable in datacenter networks, since such
networks use multi-path routing -- hence, the path an RPC takes is not
deterministic. This makes the problem combinatorial and expensive to solve.
Syslogs provide a single network-wide solution that addresses both localization
and root cause analysis.

\yperf{}'s network diagnosis module has two components: real time
problem graph mining that ingests all syslogs from a datacenter, and
asynchronous low-volume learning that periodically generates \emph{causal rules}
as input for the real time component.

\begin{figure}
\begin{center}
\includegraphics[width=.8\columnwidth]{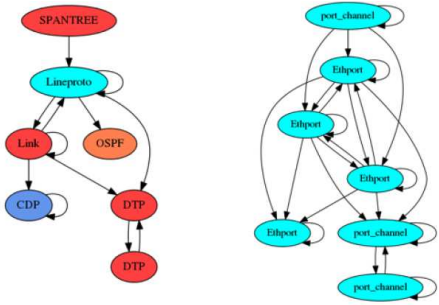}	
\caption{Example network problem graphs from our datacenter.}
\label{f:causal_graph_example}
\end{center}
\end{figure}

More formally, a problem graph is a directed graph of syslog templates,
where an edge $T_i \rightarrow T_j$ implies that template $T_i$ caused 
template $T_j$. A problem graph could exist within a single device or
span multiple devices. A causal rule connects two templates by a causal
relationship: $T_i \rightarrow T_j$; depending on whether $T_i$ and $T_j$
happen within a single device or different devices, a causal rule could be
either intra-device or inter-device. For example, Figure~\ref{f:causal_graph_example}
shows two instances of problems from a Yahoo datacenter 
(colors encode layer in protocol stack). The left graph shows
a multi-layer problem that spans aggregate and top-of-rack tiers in
the fat-tree network, and multiple layers in the protocol stack.
It encodes a cascading problem: a module failure causes a link down event, 
which triggers a spanning tree protocol status change, and causing an 
interface status change on a peering device. The right graph shows a 
problem within a top-of-rack devices that is an Ethernet (L2) flapping issue.

\yperf{}'s diagnosis module mines problem graphs as follows. 
It divides the syslog timeline across the datacenter into small time windows.
Within each time window, it maps syslog lines into templates and uses
the corpus of causal rules to iteratively construct problem graphs,
starting from intra-device edges and then adding inter-device edges.
At any point of time, we typically have 100-200 causal rules. Hence,
the runtime overhead of mining problem graphs in a small time window
of syslog messages across the datacenter is relatively low (it can run
on a single machine).

The problem of mining causal relationships between syslog templates is
relatively harder, since it is the problem of finding needles in a haystack
of syslogs. In such cases, happens-before relationships result in 
significant false positive rates. We adopt statistical causality mining
techniques to discover causal rules -- in particular, we use Quasi
Experimental Design (QED). First, we find (in a larger time window) template
pairs that have a statistically significant correlation in their timeseries.
For each template pair ${T_i,T_j}$ that is correlated, QED finds causality
by testing the hypothesis that an element of the \emph{treated} set is much more
likely than an element of the \emph{untreated} set. The treated set consists
of instances when $T_i$ and $T_j$ exists together at any time; while
the untreated set has instances when $T_i$ exists but not $T_j$.
If the treated set is more likely, QED assigns a causal relationship
$T_i \rightarrow T_j$.

For each RPC in a session that is detected as having a performance
problem, \yperf{} summarizes the set of problem graphs on devices
that the RPC could have traversed. At this point, the network diagnosis
in \yperf{} is meant to show \emph{possible} problems in the network that
impact an RPC, since these problems may not manifest as performance 
problems in all RPCs. We are working on methods to establish causal
relationship between a network problem and RPC performance. A limitation
of syslog-based diagnosis is that it will only mine problems that syslogs
can describe. We believe that our syslog-mining methodology can be
applied on logs from any multi-layer distributed service.
We refer the reader to our prior work \cite{netarxiv} for details 
of the network diagnosis methods.

%

\noindent\textbf{Internet path diagnosis.}
\yperf{} synchronously instruments the user-to-CDN (and user-to-datacenter)
path. In the context of Internet path performance, it captures userspace
RPC timing at the user host, and RPC timing at the CDN and TCP stack measurements
of the RPC at the CDN node. In practice, we observed that a common source 
of performance problems is the user host. Hence, the measurements taken from
the browser (or any container on top) include a mix of problems in the user host
and the Internet path (even after we measure and account for CDN-side latencies).
We use measurements from the TCP stack in the CDN host kernel as estimators
of the Internet path performance (as sampled by TCP). The TCP measurements
include path RTT, RTT variation, segment retransmissions, congestion windows and
reordering. \yperf{} uses the TCP measurements to estimate the impact of
the Internet path on RPCs from user host, and isolate Internet problems from
the user host performance. We are looking into using tomography on the TCP
measurements to localize bottlenecks on the Internet (in conjunction with
topology measurements).

\subsection{Ongoing Work}

\vspace{0.1in}\noindent\textbf{Process localization.} A part of our
localization goal is to simplify performance debugging by localizing
user session performance problems to source code. One approach requires
\yperf{} to track performance counters for processes within each
service and associate the counters with code; and it has to be low-overhead.
The performance counters provide a context for fingerprinting runtime
behavior of code (for node-local root cause analysis), and include 
program counters that help associate with code. This is early-stage work.

\vspace{0.1in}\noindent\textbf{Root cause analysis.} Root cause analysis
in operational practice typically relies on fingerprinting performance problems
based on domain knowledge and experience. While \yperf{} includes root cause
analysis of network problems using syslogs, an open question is how to
incorporate service and network operator inputs (domain knowledge) to
do service-level root cause analysis. The key to this is to provide a 
suitable model of performance problems that operators can input, 
using the following grammar:
\begin{lstlisting}
SYMPTOM symptom
PATHOLOGY pathology DEF ( symptom | NOT symptom )
symptom := symptom1 AND symptom2
symptom := symptom1 OR symptom2
symptom := ( symptom )
PROCEDURE symptom funcname
\end{lstlisting}
We model a performance problem as a boolean-valued expression on one or more boolean-valued
\emph{symptoms}. A symptom is a function of instrumentation.
For each detected problem, we evaluates matching problem expressions
to identify the root cause(s). The root cause analysis is based
on prior work on network root cause analysis \cite{DBLP:conf/usenix/KanuparthyD14}.

\section{YTrace Backend}

A key aspect of \yperf{} is a high-performance
analytics backend that enables near real time and accurate diagnoses. 
Figure~\ref{fig:YPerf-architecture} shows an overview of the backend.
The backend ingests YTrace instrumentation data (a timeseries of events)
and runs statistical analyses and diagnosis on the event stream. 
The events and analyses are written to a persistent store that drives
an interactive visualization system.

\vspace{0.1in}\noindent\textbf{Data transport.}
The first step after instrumentation is to transport the data to the
indexing and analysis systems. \yperf{} uses a publish-subscribe
messaging system to transport instrumentation events.

Since the \ytrace{}
libraries and the transport system implement asynchronous write semantics,
instrumentation events can incur delivery delays 
or be delivered out of order, be lost, or sometimes be duplicated.
This is particularly the case for all tracing events in a session, where
it is not always feasible to determine if all data for a session has arrived for analyses.
Moreover, due to event asynchrony, there may be statistical biases in 
certain analytics leading to false diagnosis. In our implementation, we trigger analysis
of an event after a delay $\delta$; $\delta$ is pre-computed as the 
minimum duration after which any event is delivered with a high likelihood.

In order to find biases, the \ytrace{} backend measures event volume
as a function of service and datacenter; and uses it to estimate the expected
volume at the current time. If there is a bias, it does not trigger analysis
for that statistic. Inferring and avoiding bias is a part of our ongoing work.

\vspace{0.1in}\noindent\textbf{Indexing.}
The indexing system provides a high-throughput write, low-latency read
interface for structured data. Data in \ytrace{} is a timeseries of graphs
from the network (topology) and application layers (session traces). Since
events for a session are transferred asynchronously, it is important that
the writes are idempotent and session updates do not require any reads. The
ETL process materializes a number of indices for sessions for common queries.
We currently use an Apache HBase cluster as our persistent store. Domain-specific
queries such as the network paths connecting two server hosts or Internet path
to a client host are processed by an API tier. Such queries are useful for
diagnosis such as tomography.

\vspace{0.1in}\noindent\textbf{Making it real time.}
In order to make the diagnosis near-real time, we would need to:
(1) minimize the latency between end of a session and when the events in
the session are analyzed, and
(2) build a high-throughput analytics backend.
A significant factor that contributes to the latency above is data
movement from multiple datacenters into a central indexing system
in a single datacenter. Note that on the contrary, a central 
indexing system improves the analytics throughput; however, the
latency induced by wide area data movement degrades performance
more significantly since wide area links have limited bandwidth
and are shared resources.

To reduce wide area data movement, we are working on a federated
database that is partitioned across all datacenters. Each datacenter
includes a local indexing system, and the data partitioning 
is based on the datacenter-locality of events in user sessions. 
Events inside a datacenter are transported within the datacenter; hence,
the events for a session that is served by two datacenters will reside
in two indexing systems. In the ideal case, the processing for a query
would be done at the relevant indexing systems, and the aggregated output(s)
returned to the federation layer -- the aggregations are relatively
low-volume. This is, however, not true of some queries such as joins,
which may require inter-datacenter data movement.

We are also adding support for approximate queries to speed up
query processing and reduce wide area data movement. The database
has to be aware of data biases, both from transport and partitioning,
in order to minimize statistical bias in query output. Our work builds
on prior work in wide area and approximate query processing (e.g., the
recent work on WANalytics \cite{vulimiri2015wanalytics} and BlinkDB \cite{agarwal2014knowing}).

\section{Case Studies}

In this section, we show some experiences and results using \yperf{}
in production.

\begin{figure}
\begin{centering}
\includegraphics[angle=270,width=1\columnwidth]{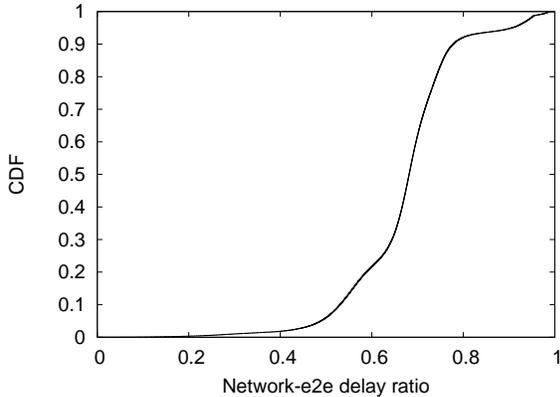}
\par\end{centering}

\protect\caption{Effect of network delays on about 2m user operations with the \sherpa{}
data store from two \yahoo{} datacenters.\label{fig:Results-sherpa}}
\end{figure}

\subsection{Distributed Storage}

We consider a hosted large-scale, low-latency, distributed key-value 
storage system, \sherpa{}~\hide{\cite{cooper2008pnuts}}{}, that
is used as a common storage backend in serving stacks at Yahoo in Figure
\ref{fig:Results-sherpa}. \sherpa{} aims for an SLO of 2ms for key reads.
We look at a multi-layer analysis of latencies in \sherpa{}.

Operations with \sherpa{} traverse router nodes and are served by Storage Unit (SU) nodes --
all connected by the datacenter network. We use \ytrace{} data to
look at the impact of round-trip network latencies\footnote{Latency computations 
in this study use a single clock. Network latency is: (router-SU
RPC exchange at router) - (processing delay at SU).} on latency of
two million \sherpa{} operations in two datacenters. The figure shows that the network
contributes to a significant fraction of operation latency. The tail
of the distribution (top-10\%) includes operations that saw variable
delays in the network (e.g., due to congestion or non-shortest path 
routing).

Using a simple model of network delay for a key read payload, we
can show that the minimum delay for a read RPC to traverse the router
and SU nodes and back is 0.5 to 0.9ms (depending on the number of round-trips
TCP takes). Under per-hop queueing or non-shortest path routing conditions
(a router-SU path normally traverses 1-2 ToR and/or one AGG device), the
delay can be 0.7-1.3ms. Hence, in order to optimize for operation latency
and maintain SLOs, the storage system could be designed to minimize the number of 
network hops traversed by RPCs.

\subsection{Datacenter Network}

We look at datacenter-wide problems from a single Yahoo datacenter
using \yperf{}'s network diagnosis output. The datacenter consists of
a large fat-tree network topology with thousands of network devices.
The topology is made of multiple ``tiers'': traversing bottom-up,
Top-of-Rack (ToR) devices that connect servers (running services),
multiple aggregation (AGG) tiers and a core tier that connects the datacenter
to the Internet. RPCs between services within the datacenter typically
traverse the ToR and AGG tiers; hence, any problems in the two tiers
will impact a significant fraction of RPCs in the datacenter.

\begin{figure}
\begin{center}
\includegraphics[width=1\columnwidth]{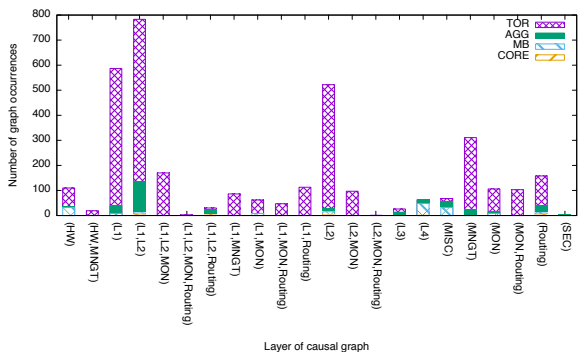}
\caption{Number of problem graphs for each network tier.}
\label{f:gsig_cnts_layer}
\end{center}
\end{figure}

Figure~\ref{f:causal_graph_example} shows two examples of problem
graphs from ToR and AGG tiers in the datacenter network; see \S\ref{sec:netdiag}
for details. Figure~\ref{f:gsig_cnts_layer} shows the distribution of
different problem classes across the three network tiers. Over 93\% of
the problems occur in the ToR switches (which dominate in number and are
relatively low-cost devices). A large fraction of ToR and AGG problems occur
in the lower layers (PHY and L2), and sometimes in higher layers such as
the routing plane. On the other hand, middleboxes (that can be topologically placed anywhere
in the network) show problems mostly in the higher layers (L3 and L4). The
duration of the problem graphs can last between a few seconds to hundreds
of seconds -- which makes the likelihood of RPCs being affected high. 
We refer the reader to our prior work \cite{netarxiv} for more results and
details of network diagnosis methods.


\begin{figure}[t]
\begin{centering}
\includegraphics[angle=270,width=1\columnwidth]{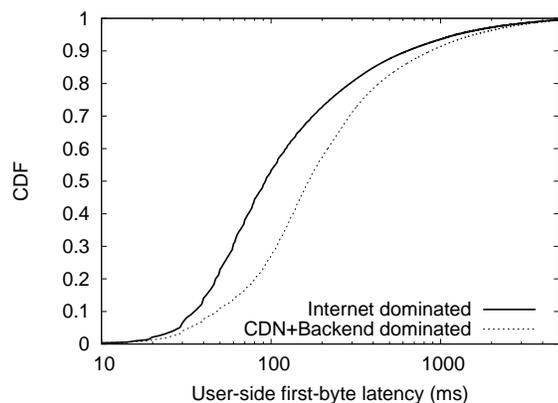}
\end{centering}
\caption{User-side RPC first-byte latency broken down by the dominant bottleneck (Internet or CDN).\label{fig:cdn-fb}}
\end{figure}

\begin{figure}
\begin{centering}
\includegraphics[angle=270,width=1\columnwidth]{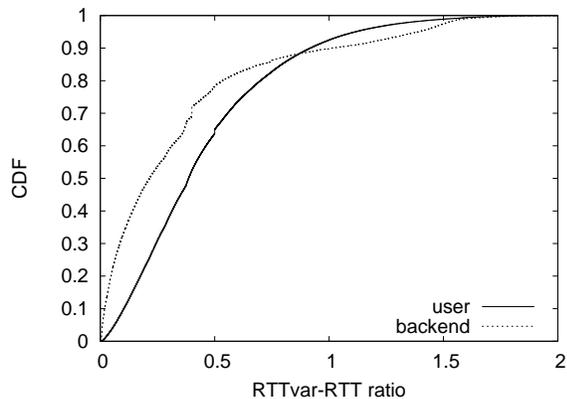}
\par\end{centering}
\protect\caption{Network RTT variation of user and datacenter paths in 350m user sessions
with the \yahoo{} CDN.\label{fig:Results-cdn}}
\end{figure}

\subsection{Video Stack}

Video serving stacks can be modeled as three tier architectures, spanning
the video player (user-side), the CDN and a backend store. A video
playback is a sequence of RPCs by the player to the CDN; the CDN makes an 
RPC to the backend store if it does not have the response cached.
We trace RPCs from the video player through the CDN, while synchronously
instrumenting the TCP stack in the CDN kernel periodically over the
course of the RPC. We use the TCP measurements as the source of truth
for the Internet path performance, since the delays induced by the kernel
space are relatively very low. We collect data for all user sessions 
over a course of two weeks for this case study. 

We first look at the impact of backend and Internet performance on the
user experience. We quantify the user experience as the first-byte delay
for each RPC. Figure~\ref{fig:cdn-fb} shows the distribution of user-side
latencies after dividing the set of RPCs into two parts: RPCs that are
bottlenecked by the Internet path and RPCs that are bottlenecked by the
CDN or backend. About 95\% of the RPCs are bottlenecked by the Internet
path as we would expect. In the remaining 5\% RPCs that are bottlenecked
by the CDN/backend, the cache miss rate is 40\% as compared to 2\% overall.
Further, the user experience degradation due to RPCs bottlenecked by
the CDN/backend is tens of milliseconds higher than RPCs bottlenecked
by the Internet. This shows that in order to troubleshoot or fix tail
latencies, we should focus on the CDN/backend.

\begin{figure}
\begin{centering}
\includegraphics[angle=270,width=1\columnwidth]{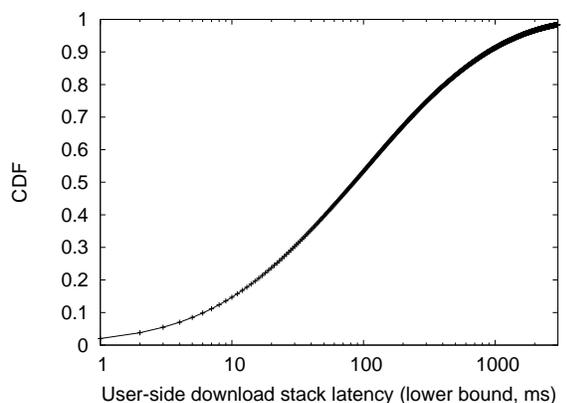}
\par\end{centering}
\protect\caption{User-side stack latency estimates (lower bound).\label{fig:dds}}
\end{figure}

When there is a cache miss, the CDN makes an RPC to the backend.
We look at RTT of TCP connections at the \yahoo{} CDN -- for RPCs from the user
host and to the backend in Figure \ref{fig:Results-cdn}. The
RTT is the delay between a TCP data segment and the corresponding
TCP acknowledgement at the CDN node's TCP stack. For 350m user sessions,
we analyze the \emph{variation} in RTT of each TCP connection, defined
as the ratio of RTTvar and smoothed RTT in the kernel.
We see that the RTT variation is significantly
higher for user connections than datacenter connections; however,
the tail RTT variation is dominated by the datacenter connections.
This also makes the case for troubleshooting tail latency problems
by looking at the CDN/backend.

Finally, we show that it is possible to estimate content download
latencies in the user host by tracing instrumentation alone (i.e.,
timing at user and CDN hosts, and TCP kernel variables). The first-byte
delay at the user host for video segment RPCs to the CDN ($\delta_{fb}$)
includes an RTT ($\delta_{rtt}$) on the Internet, CDN and backend
(if any) latencies ($\delta_{cdn}$ and $\delta_{be}$), and the download
stack latency at the user host ($\delta_{ds}$). Considering the TCP Retransmission Timeout ($\delta_{rto}$)
as a conservative estimate for $\delta_{rtt}$, we can estimate a lower
bound for $\delta_{ds}$: $\delta_{ds} \ge \delta_{fb} - \delta_{cdn} - \delta_{be}- \delta_{rto}$.
Figure~\ref{fig:dds} shows the distribution of the lower bound of 
download stack latencies across video sessions (truncated to 
positive real numbers). We see that the user host contributes tens to
hundreds of milliseconds of latency when delivering data to the 
video player application running on the browser. Although user-side
download stack delays are not under the control of a content provider,
providers can avoid the effect of such problems by ordering
RPCs to mask the problem. We refer the reader to our prior work \cite{2016arXiv160504966G}
for more results on the video stack.

\section{Related Work}

Diagnosis systems are typically designed for diagnosing a subset
of the end-to-end path or a specific layer of the stack. 
\yperf{} is an attempt to build an end-to-end
multi-layer diagnosis system at web-scale, since performance troubleshooting
activities typically rely on such insights. In doing so, it 
builds on some prior systems. We capture representative work in this section.

\noindent\textbf{Distributed tracing systems.}
Distributed tracing is a common instrumentation primitive in content
providers. Capturing, recording or mining causality between events in a distributed
trace is necessary to make sense of session performance. Systems in
prior work differ in the amount of instrumentation and trace analysis 
complexity -- in fact, there is a tradeoff between instrumentation overhead
and analysis complexity to do the same amount of diagnosis.

Systems implement causality synchronous with execution \cite{fonseca2007x,sigelman2010dapper}
or mine using historic traces \cite{chow2014mystery}. For example,
Dapper (and its derivative, Zipkin) capture causality between \emph{spans},
which are combinations of requests and associated responses. While
span-level causality is useful, it is not expressive enough to model
RPC executions such as parallelism and redundancy. History-based
causality mining helps minimize instrumentation overhead in production;
it relies, however, on resources for offline mining of causal relationships.
It works well in homogeneous environments, where there is a common RPC
library and RPC execution patterns are predictable, but 
may not be feasible in heterogeneous and dynamic runtime
environments due to runtime transitions to non-causality within a
session. Magpie \cite{barham2004using} lies on the instrumentation side of
the spectrum -- it captures detailed instrumentation, such as OS events
and packet traces, to infer causality without needing offline analysis.
While this yields detailed diagnoses, it may not be feasible in production.
Project 5 \cite{aguilera2003performance}, Mystery Machine \cite{chow2014mystery} 
and Pinpoint \cite{} lie on the analysis side of the spectrum -- they require
offline resources to mine causality.
X-Trace is an experimental system that requires session 
tracing support from network devices; having such support helps do
multi-layer causal discovery synchronously with the session (a limitation
of \yperf{}), but network support may not be feasible in practice.

\noindent\textbf{Network diagnosis.}
There has been significant research on network diagnosis methods. Sun et al.
capture TCP variables at the CDN to localize performance bottlenecks
\cite{sun2011identifying}; they require OS kernel changes in the
critical path, since they require TCP instrumentation outside of the
\texttt{tcp\_info} structure. WhyHigh \cite{krishnan2009moving} and LatLong \cite{zhu2012latlong}
further discover client clusters with performance problems and diagnose
user-to-CDN path problems at an aggregate level (instead of per-session). 
Yu et al. \cite{yu2011profiling} and Ghasemi
et al. \cite{Ghasemi:2013:RDT:2537148.2537156} diagnose datacenter
network performance using detailed instrumentation (e.g.,
socket logs and packet traces). At large serving rates, such logging
may be infeasible. Network tomography techniques
\cite{Duffield:2003:SNP:948205.948232} localize bad performance to 
network interfaces; they assume that the path between two hosts is
known -- uncommon in datacenter networks.
MonitorRank \cite{kim2013root} uses similar tomography-based localization.

\noindent\textbf{Log mining.}
Service and network log mining are common diagnosis methods.
Distalyzer compares anomalous logs with known baseline
logs \cite{nagaraj2012structured} for diagnosis. 
Spectroscope compares two trace logs to understand
differences between them \cite{sambasivan2011diagnosing}. 
Xu et al. mine log features \cite{xu2009detecting}. 
Syslogs have been used to study network-specific failures in datacenters~\cite{Gill:2011:UNF:2018436.2018477,Potharaju:2013:NCE:2523616.2523638,Potharaju:2013:DDS:2504730.2504737},
but not for root cause analysis. Prior work has not
explored causal discovery for log analysis -- this becomes
particularly necessary when looking for a small number of
cascading problems in large log volumes.

\noindent\textbf{Code and content localization.}
Binary profiling \cite{anderson1997continuous,attariyan2012x,ren2010google}
and code mining methods \cite{186221} have been used to diagnose
performance problems in single hosts down to code. These systems
do not track code-level problems with user experience.
More recently, Pivot Tracing \cite{mace2015pivot} allows users
to insert breakpoints in running code and log them while tracing
(synchronously) -- mainly tailored towards debugging within a distributed
system (as opposed to a content provider). Content diagnosis methods used browser
modifications \cite{wang2013demystifying} and middleboxes \cite{Kiciman:2007:APR:1294261.1294264}.
We are exploring the feasibility of these methods in \yperf{}.

\section{Discussion and Conclusion}\label{sec:Discussion-and-Conclusion}

In this paper, we presented the design of \yperf{}, a system for
end-to-end multi-layer performance diagnosis in large content providers.
We formulated a problem statement that covers diagnosis use cases, and
presented instrumentation and methods for diagnosis.
Our discussion opens several research questions that we cover next.

If an RPC is observed to have high latency in the datacenter, 
\yperf{} currently lists correlated network problems 
from devices that the RPC potentially traversed (based on syslogs).
The longer the problem, the more likely that RPCs that traversed the
devices will be impacted. Such correlations are useful towards
troubleshooting (esp. when looking at aggregated data).
Going from correlation to causation -- in other
words, whether a network problem \emph{caused} performance problems with
RPCs -- is an open question. It requires apriori knowledge of 
devices the RPC traversed, e.g.,  using Netflow (since datacenter networks use multi-path routing),
and inferring causal relationships between problems in those devices
and RPC performance. One approach is to look for symptoms in RPCs
instrumentation that are caused by each network pathology.

\yperf{} diagnoses distributed root causes such as cascades in the datacenter
network, but does not diagnose distributed root causes across services.
Such problems create runtime dependencies between services that impact
performance (despite RPC parallelism and redundancy).
A common service-level cascading problem is backlog that builds up across
services (typically calling services). Distinguishing these from backlogs
that arise due to problems within the host requires appropriate
instrumentation and diagnosis methods. We are looking into adopting
causality-based joint mining of service logs and network syslogs to
diagnose distributed root causes.

We assumed that Traffic Engineering (TE) at the CDN is a given: \yperf{}'s
diagnosis is conditioned on the TE for a user session. Diagnosing
performance\emph{-sub-optimal} TE for a session (i.e., whether a user
was directed to a CDN node that caused bad user experience) requires
knowledge of Internet path performance from the user to all CDN nodes
at that time; accurately doing it is an open research direction. A
related system, LatLong, diagnoses \emph{average} latency \cite{zhu2012latlong}.

Content providers may not have a complete view of the user end-host
stack performance (hardware, OS environment, browser, etc.) as the
content is parsed, executed and rendered. Analysis similar to WProf
\cite{wang2013demystifying} that does not require browser modifications
would help diagnose bottlenecks that reside in the user end-host,
and could be exposed to the content provider similar to Navigation
Timing \cite{navtiming}.

In order to reduce overhead due to instrumentation, \yperf{}
supports sampling a fraction of user sessions. Sampling leads to challenges
in analyzing tail latency -- it requires inversion of the 
sampled distribution of execution graphs to estimate high quantiles. For example,
prior work on latency looked at estimating confidence intervals for
latency quantiles \cite{sommers2010multiobjective}.

In order to localize performance problems to networks and inter-domain 
links on the Internet, we are looking into adopting tomography methods that
work on TCP measurement data. Tomography methods
assume that the Internet path for a user IP address is known.
In practice, provider networks may use multipath routing. Without additional active probing
or data (e.g., Netflow) at the time of the RPC flow \cite{pan2014end},
it is challenging to find the sequence of Internet hops that a given RPC took.

Finally, Internet and transit providers can deploy traffic management
mechanisms that do not follow conventional wisdom and can impact 
performance. For example, traffic shaping leads to changes in link 
capacity, which can impact long-running
flows. \yperf{} currently diagnoses such mechanisms as a part of
the user-CDN Internet path; diagnosing such root causes, however,
is an open problem. Recent work on tomography shows that the methods
can be used (under sufficient sample size) to find content discrimination
\cite{Zhang:2014:NNI:2740070.2626308}, under assumptions of static routing.

Web-scale performance diagnosis requires re-thinking from ground up:
the instrumentation design, algorithms and systems design to enable
near real time diagnoses. There is an inherent tradeoff between complexity
of and how detailed diagnosis could be, versus the amount of per-session
instrumentation volume we can collect in production at scale. Traditional
methods such as tomography and blackbox RPC causality learning are hard to
apply in large-scale heterogeneous cloud environments. \yperf{} is an attempt to accomplish
performance diagnosis at scale.

\begin{small}
\section*{Acknowledgements}
Ahmed Mansy led the TCP instrumentation effort.
We would like to thank many Yahoos who we had insightful
discussions with, and who contributed to YTrace instrumentation, deployment
and backend: Jayanth Vijayaraghavan, Vahid Fatourehchi, Joshua Blatt, Christophe Doritis,
Vidya Srinivasan, Srikanth Sampath Kumar, Arvind
Murthy, Nishant Mishra, Jacob Cherackal, Powell Molleti, Antonia Kwok,
Bhupendra Singh, Seema Datar, Evan Torrie, Rupesh Chhatrapati, Maurice
Barnum, Amit Jain, Ramachandran Subramaniam,
Amar Kamat, Tague Griffith, Archie Russell, Tim Miller, Scott Beardsley,
Amotz Maimon, Ian Flint, Rick Hawes and Benoit Schillings.
We thank Chen Liang (Duke) for syslog analysis and Jennifer Rexford 
(Princeton) for helpful discussions on the video stack.
\end{small}

\balance

{\footnotesize{}\bibliographystyle{acm}
\bibliography{ref}
}{\footnotesize \par}
\end{document}